\documentstyle[amssymb,aps,prb,twocolumn]{revtex}

\newcommand{\be}{\begin{eqnarray}}
\newcommand{\ee}{\end{eqnarray}}

\begin{document}

\twocolumn[\hsize\textwidth\columnwidth\hsize\csname @twocolumnfalse\endcsname

\title{Structural and superconducting properties of MgB$_{2-x}$Be$_x$}
\author{J. S. Ahn$^{1}$, Young-Jin Kim$^{2}$, M.-S. Kim$^3$, S.-I. Lee$^3$, and E.
J. Choi$^{1,2}$}
\address{$^1$Center for Strongly Correlated Materials Research, Seoul National\\
University, Seoul 151-742, Republic of Korea}
\address{$^2$Department of Physics, University of Seoul, Seoul 130-743, Republic of\\
Korea}
\address{$^3$Center for Superconductivity and Department of Physics, \\
Pohang University of Science and Technology,\\
Pohang 790-784, Republic of Korea}
\maketitle

\begin{abstract}
We prepared MgB$_{2-x}$Be$_{x}$ ($x=0$, 0.2, 0.3, 0.4, and 0.6) samples
where B is substituted with Be. MgB$_{2}$ structure is maintained up to $%
x=0.6$. In-plane and inter-plane lattice constants were found to decrease
and increase, respectively. Superconducting transition temperature $T_{c}$
decreases with $x$. We found that the $T_{c}$ decrease is correlated with
in-plane contraction but is insensitive to carrier doping, which is
consistent with other substitution studies such as Mg$_{1-x}$Al$_{x}$B$_{2}$
and MgB$_{2-x}$C$_{x}$. Implication of this work is discussed in terms of
the 2D nature of $\sigma $-band.
\end{abstract}

\date{\today }

\vskip2pc]


\begin{section}{Introduction}%
%

The recent discovery of superconductivity in MgB$_{2}$ at $T_{c}=39$ K has
drawn enormous attention in its structural and electrical properties.\cite
{Akimitsu2001} The borons form the graphitic planes and magnesiums supply
charges in the planes. The resulting carriers are holes in the $\sigma $%
-band.\cite{Kang2001} Most theoretical works suggest that coupling of the $%
\sigma $-hole with B-plane phonon is the key ingredient of the
superconductivity in this compound.\cite{Kortus2001,An2001,Liu2001} This
phonon mediated BCS mechanism is consistent with the boron isotope effect. 
\cite{Budko2001}

There have been many attempts to change $T_{c}$ through chemical
substitution in Mg- or B-sites. For example, Al-substitution on Mg-site\cite
{Slusky2001} and C-substitution on the boron plane are reported.\cite
{Ahn2001} Such chemical substitutions change physical quantities of the
system such as hole density, lattice constants, etc. However, there is no
detailed understanding on the effect of substitution on the observed $T_{c}$
change even within the BCS frame.

In this paper, we prepared a series of MgB$_{2-x}$Be$_{x}$, where B is
substituted with Be in the plane.\cite{Akimitsu2} With this substitution,
the in-plane B-B distance decreases while the inter-plane distance
increases. Our thermopower measurement showed that $\sigma $-hole increases
with Be, as reported in the independent paper.\cite{Ahn2001TEP} We compare
our result with other substituted compounds and find that the $T_{c}$ change
is insensitive to the carrier doping. Instead, we show that the in-plane B-B
distance is closely correlated with the $T_{c}$ change in the low doping
region. This result is consistent with the 2D nature of the $\sigma $-band. 
\cite{An2001}

\end{section}%
%

\begin{section}{Experiment}%
%

Polycrystalline samples were synthesized with the powder metallurgical
technique under high pressure.\ Starting materials were fine powders of Mg
(99.8\%, Alfa Aesar), amorphous B (99.99\%, Alfa Aesar), and Be (99.+\%,
Alfa Aesar). Stoichiometric amounts of powders were mixed and pelletized.
The pellets were placed in a tungsten vessel with a close-fitting cap, then
reacted for two hours at 850 $^{\circ }$C under 20 atm. of high purity argon
atmosphere.

X-ray diffraction (XRD) $\theta -2\theta $ scan measurement was performed
using a Rigaku RINT d-max. Figure 1(a) shows the results for $x=0$, 0.2,
0.4, 0.6, and 1.0. Most of the reflections correspond to the AlB$_{2}$-type
patterns. Also, minor impurity phases such as MgO and BeO are found as
indicated by $\nabla $ and *, respectively. Note that the MgB$_{2}$
structure is maintained up to $x=0.6$. At higher compositions ($x=1.4$ and
2.0, not shown here), we find that Be$_{13}$Mg becomes the main phase. In
Fig. 1(b), the shifts of (002) and (110) reflections are shown in expanded
scales. As $x$ increases, (002) reflection shifts to the lower angle while
(110) reflection moves to the opposite direction. This indicates that the
in-plane lattice constant $a$ decreases and the inter-plane distance $c$
increases with $x$.\cite{Mehl2001} To obtain the lattice parameters, we
performed refinement analysis using the RIETAN-2000 program\cite{Izumi} and
used MgO and BeO as internal standards.

In order to study superconducting property, dc-magnetization was measured
using a dc SQUID magnetometer (Quantum Design). Figure 2 shows magnetization 
$M(T)$ of MgB$_{2-x}$Be$_{x}$. In this measurement, samples were first
zero-field-cooled (ZFC) and data were measured with increasing temperature
under $H=10$ Oe. Note that the superconducting transition temperature $T_{c}$
decreases with $x$. In the figure, the magnetization is normalized with the
saturated value at $T=5$ K. The transition width $\Delta T_{c}$ was
determined from 10 $\sim $ 90 \% the transition.

\end{section}%
%

\begin{section}{Result and Discussion}%
%

In figure 3, we summarize the XRD and $T_{c}$ results. The upper panel shows
the lattice constants $a$ and $c$. $T_{c}$ and $\Delta T_{c}$ are shown in
the lower panel. With Be-doping, $a$ decreases while $c$ increases. Note
that the change in $c$ ($\sim $ $3.3$ \%) is much larger than that in $a$ ($%
\sim 0.7$ \%). The transition temperature $T_{c}$ decreases but $\Delta T_{c}
$ shows maximum near $x=0.3$. This is an interesting result, because
normally the superconducting transition becomes broader with random
substitutions. We associate this anomalous behavior of $\Delta T_{c}$ with
the structural data: we note that the (002) reflection is the most broad at $%
x=0.3$ in Fig. 1(b), which indicates considerable distribution in $c$ value,
possibly due to the two stable phases at $x=0$ and $x=0.6$. This effect may
increase $\Delta T_{c}$ at the intermediate $x$ values. Considering the
drastic change of $T_{c}$, the tansition width will be more meaningful when
it is normalized with $T_{c}$. As shown with open squares, the ratio $\Delta
T_{c}/T_{c}$ increases with $x$.

Now let us consider the $T_{c}$ change. To see a possible correlation of $%
T_{c}$ with lattice parameters, we first plot $T_{c}$ vs. $c$ in Fig. 4(a).
Our results for MgB$_{2-x}$Be$_{x}$ are shown as solid circles. For
comparison, the results for Mg$_{1-x}$Al$_{x}$B$_{2}$ (open squares)\cite
{Slusky2001} and MgB$_{2-x}$C$_{x}$ (open circles)\cite{Ahn2001} are also
shown. In MgB$_{2-x}$Be$_{x}$, $c$ expands with the substitution while it
contracts in Mg$_{1-x}$Al$_{x}$B$_{2}$. For MgB$_{2-x}$C$_{x}$, $c$ remains
almost unchanged. For all these three different cases, $T_{c}$ decreases and
there is no correlation between $T_{c}$ and $c$. This indicates that the
inter-layer distance is not directly related with $T_{c}$.

Next, we examine $T_{c}$ vs. $a$ relation in Fig. 4(b). For all the samples, 
$a$ shrinks with the substitutions. Along with the $a$ decrease, $T_{c}$
decreases linearly. With further contraction, sudden drops are\ found for MgB%
$_{2-x}$Be$_{x}$ and Mg$_{1-x}$Al$_{x}$B$_{2}$. This indicates some abrupt
changes in the samples, for example structural instability like buckling or
ordering in the plane. It is useful to compare the linear $T_{c}$ decrease
with the result of high pressure experiment on pristine MgB$_{2}$ (dashed
line): As external pressure is applied, the lattice is compressed and $T_{c}$
is found to decrease.\cite{pressure} (The dashed line in Fig. 4(a)
represents the pressure result for $c$ compression) Note that the $T_{c}$
decrease in substituted compounds is similar to the pressure line when the
contraction ($-\Delta a/a_{0}$) is small.

In Fig. 4(c), we examine the low contraction region more closely. Note that
the data of the two compounds (Be-, and Al-substituted) fall nearly on the
same line (see the solid guide line). It is interesting that the $T_{c}$
change occurs in the same direction (decrease) for both electron (Al-\cite
{Lorenz2001TEP}) and hole (Be- \cite{Ahn2001TEP}) doping. In fact, such a
behavior is predicted theoretically by An et al. \cite{An2001}: their band
calculation for MgB$_{2}$ shows that the in-plane $\sigma $-band is highly
2D like and the DOS is nearly constant with energy. In this case, the Fermi
level shift due to carrier doping will not affect $T_{c}$. Our observation
is consistent with this prediction.

Our results suggest that the $a$ contraction plays the key role in the
observed $T_{c}$ decrease.  According to the BCS analysis of MgB$_{2}$, \cite
{Loa2001} the shrinkage in $a$ leads to decrease in the DOS, increase in the
phonon frequency, and decrease in the electron-phonon coupling. As a result, 
$T_{c}$ decreases.\cite{Lorenz2001,Goncharov2001,Vogt2001} Note that for Al-
and Be-substituted samples, $T_{c}$ decreases somewhat faster than the
pressure line, suggesting that some additional effects such as, for example,
structural randomness arise by the substitution. It is interesting that in
MgB$_{2-x}$C$_{x},$ the slope is close to the pressure line, suggesting that
the additional effects are minimal.\cite{above}. Our observation points to
possible $T_{c}$ increase if $a$ could be expanded.\cite{Zn} In fact,
Medvedeva et al. predicted that higher $T_{c}$ may be obtained in lattice
expanded case such as the hypothetical CaB$_{2}$.\cite{Freeman2001}

It is interesting to compare our results with the theoretical prediction on
MgB$_{2-x}$Be$_{x}$ by Mehl et al.\cite{Mehl2001}. According to their
calculation, the lattice constant $a$ increases with $x$, opposite to our
observation. At this point, it is not clear where the discrepancy arise
from. As one possibility, we speculate that Be substituted in the B plane
may actually exist in partially ionic state rather than in prefect covalent
state. In this case, Be radius is smaller than its covalent radius, which
can result in the in-plane contraction.\cite{Pearson}

\end{section}%
%

\begin{section}{Conclusion}%
%

In this work, we studied structural and superconducting properties of MgB$%
_{2-x}$Be$_{x}$ ($x=0$, 0.2, 0.3, 0.4, and 0.6). In-plane and inter-plane
lattice constants were found to decrease and increase respectively with
Be-substitution. While $T_{c}$ decreases with $x$ monotonically, $\Delta
T_{c}$ shows maximum at $x=0.3$. From our results and other substitution
studies (Al- and C-substitution), we found that the $T_{c}$ change is
correlated with the in-plane contraction and is independent of the carrier
doping. This is consistent with the 2D nature of the $\sigma $-band.

\end{section}%
%

\begin{section}*{Ackowledgments}%
%

This work was supported by KRF-99-041-D00185 and by the KOSEF through the
CSCMR. Work at POSTECH was supported by the Creative Research Initiatives of
the Korean Ministry of Science and Technology.

\end{section}%
%

\begin{figure}[tbph]
\caption{(a) x-ray $\protect\theta -2\protect\theta $ scan results of MgB$%
_{2-x}$Be$_{x}$ ($x=0$, 0.2, 0.3, 0.4, 0.6, and 1.0). Impurity peaks are
indicated with symbols: MgO($\protect\nabla $) and BeO(*). (b) Shift of
(002) and (110) reflections.}
\label{Fig:1}
\end{figure}

\begin{figure}[tbph]
\caption{Magnetization $M(T)$ of MgB$_{2-x}$Be$_{x}$. ZFC results are shown.
External magnetic field $H=10$ Oe is used. Data is normalized with 5 K
value. }
\label{Fig:2}
\end{figure}
\begin{figure}[tbph]
\caption{Upper: lattice parameters $a$ ($\bullet $) and $c$ ($\bigcirc $)
vs. $x$. Lower: superconducting onset temperature $T_{c}$ ($\blacktriangle $%
), transition width $\Delta T_{c}$ ($\triangle $), and the ratio $\Delta
T_{c}/T_{c}$ ($\square $) vs. $x$. $\Delta T_{c}$ is determined from the 10 $%
\sim $ 90 $\%$ transition of saturated magnetization.}
\label{Fig:3}
\end{figure}

\begin{figure}[tbp]
\caption{$T_{c}$ change with respect to lattice compression for various
chemically substituted compounds. (a) $T_{c}$ vs. inter-plane lattice
contraction $-\Delta c/c_{0}$. (b) $T_{c}$ vs. in-plane lattice contraction $%
-\Delta a/a_{0}$. (c) shows expanded view of low doping region of (b).
Experimental data are shown with symbols: MgB$_{2-x}$Be$_{x}$ ($\bullet $),
Mg$_{1-x}$Al$_{x}$B$_{2}$ ($\square $), and MgB$_{2-x}$C$_{x}$ ($\bigcirc $%
). Results of high pressure measurement for pristine MgB$_{2}$ are shown for
comparison (dashed lines), based on the experimental results below 1 GPa. 
\protect\cite{pressure} Solid and dash-dotted lines are for eye-guide.}
\label{Fig:4}
\end{figure}


\begin{references}
\bibitem{above}  Note that the pressure line lies above the other
substituted data. $T_{c}$ of the pressure line is taken from the resistivity
data while the others are taken from the magnetic measurement. Usually the
resistivity gives a little bit higher $T_{c}$ in substituted samples of MgB$%
_{2}$.
\end{references}
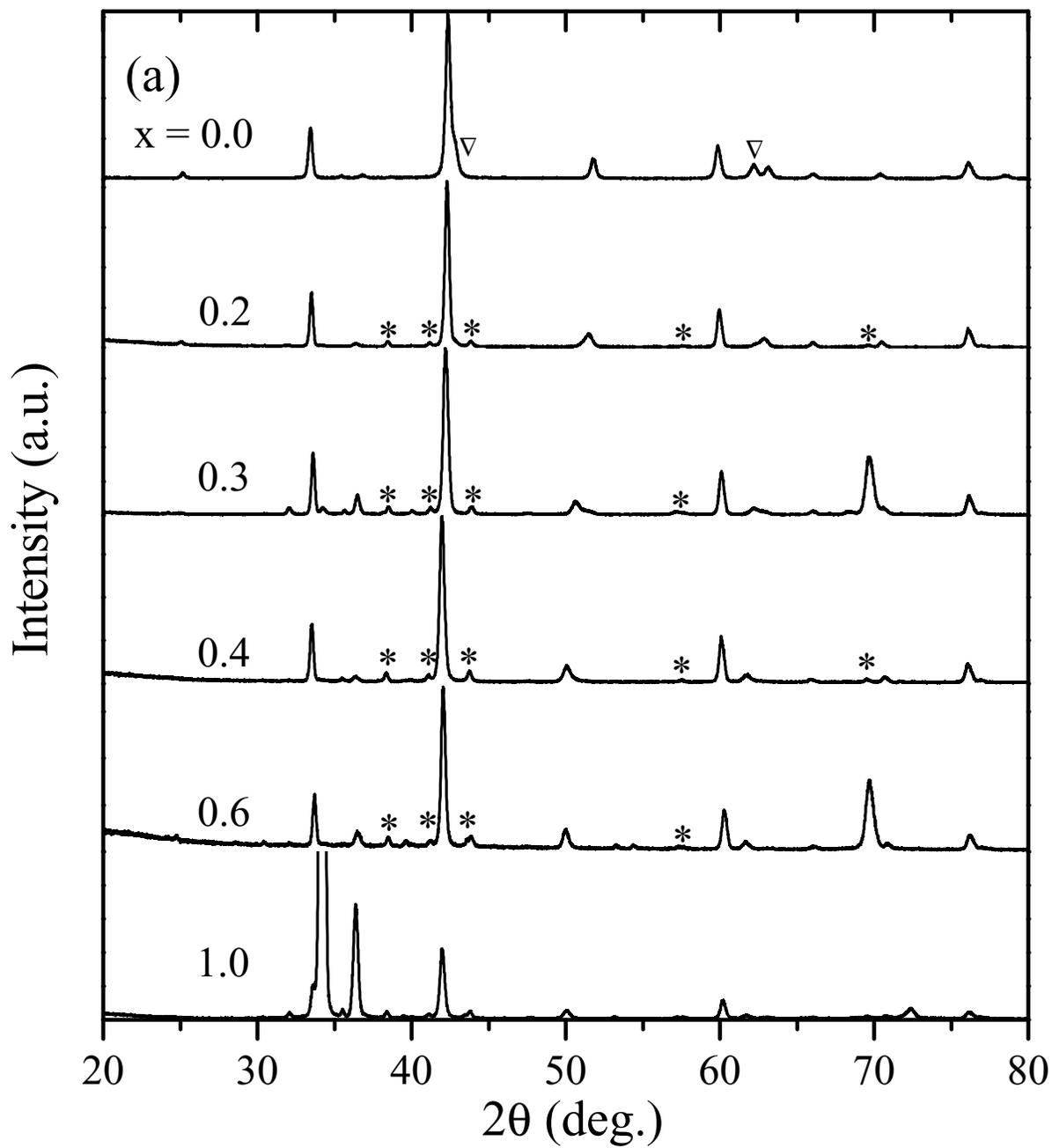
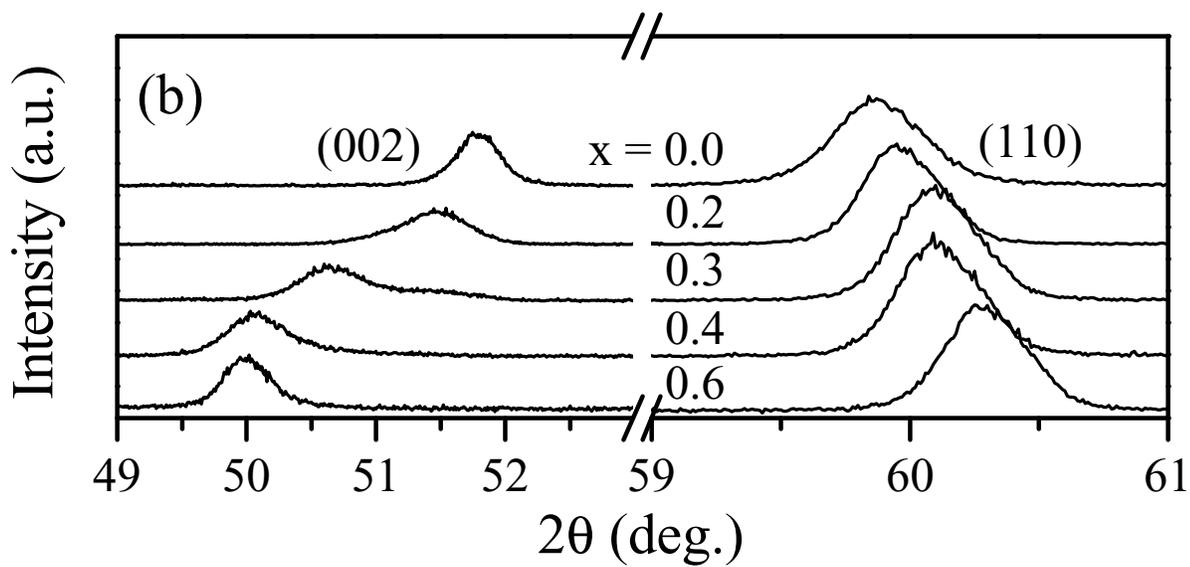
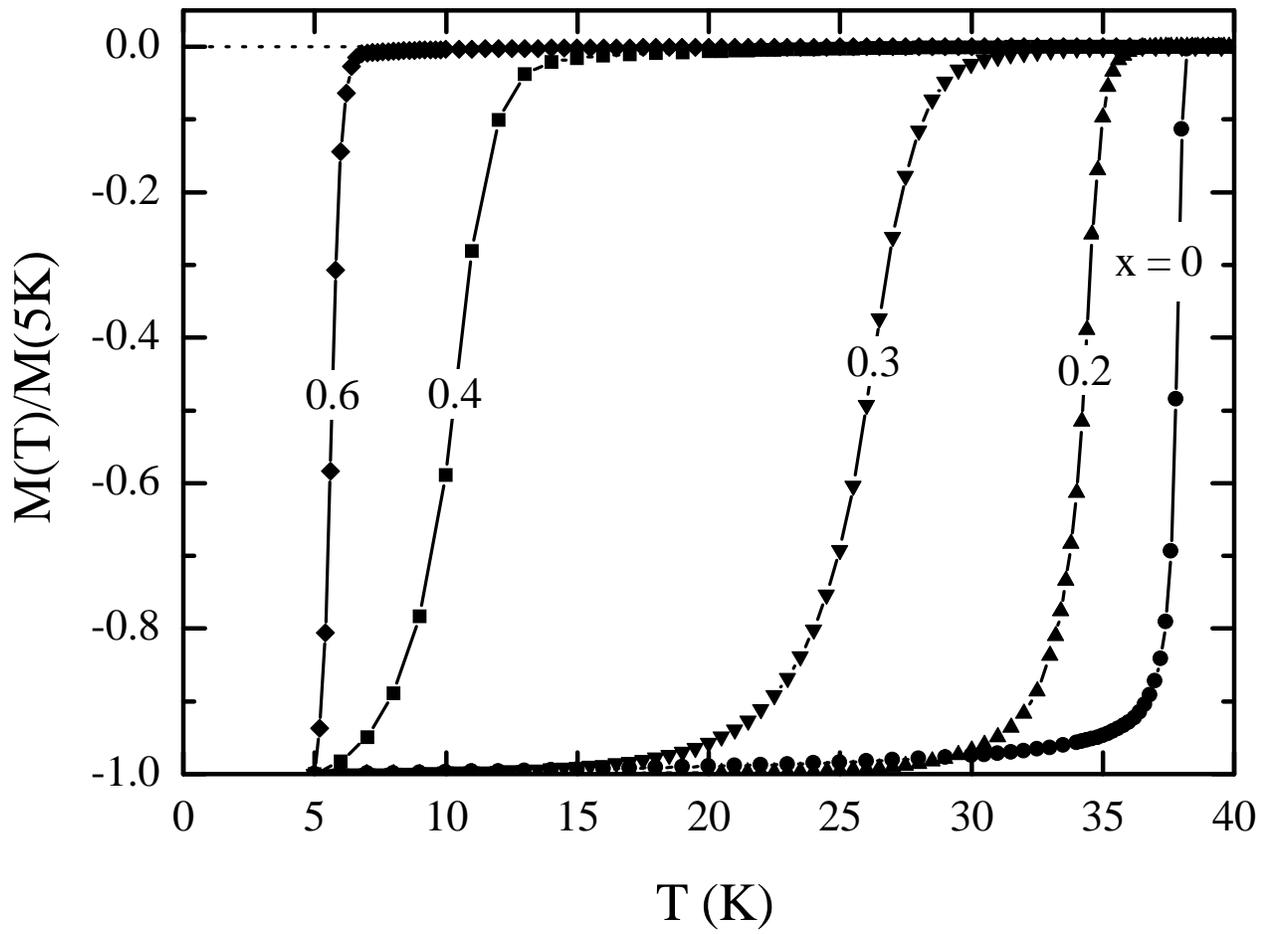
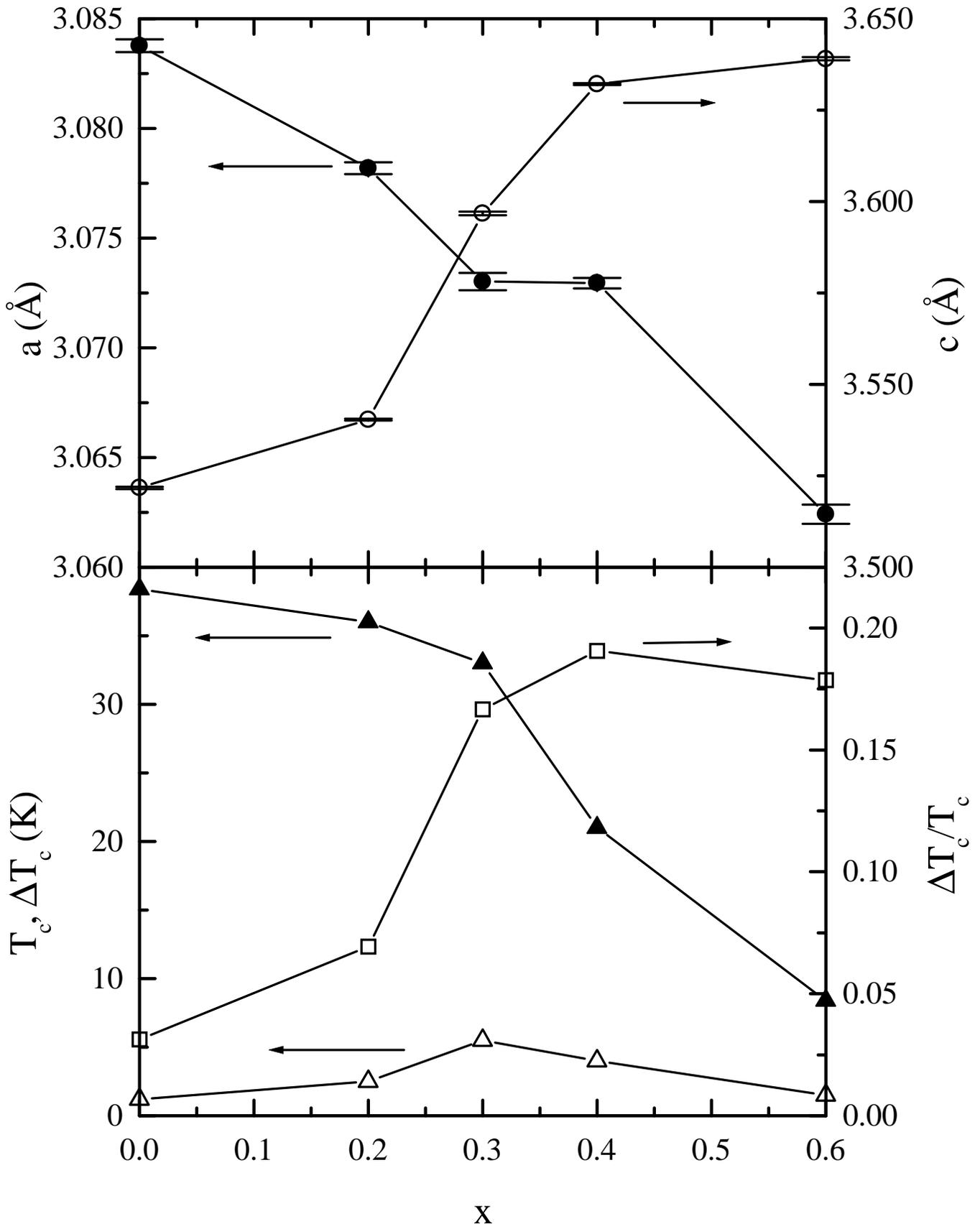
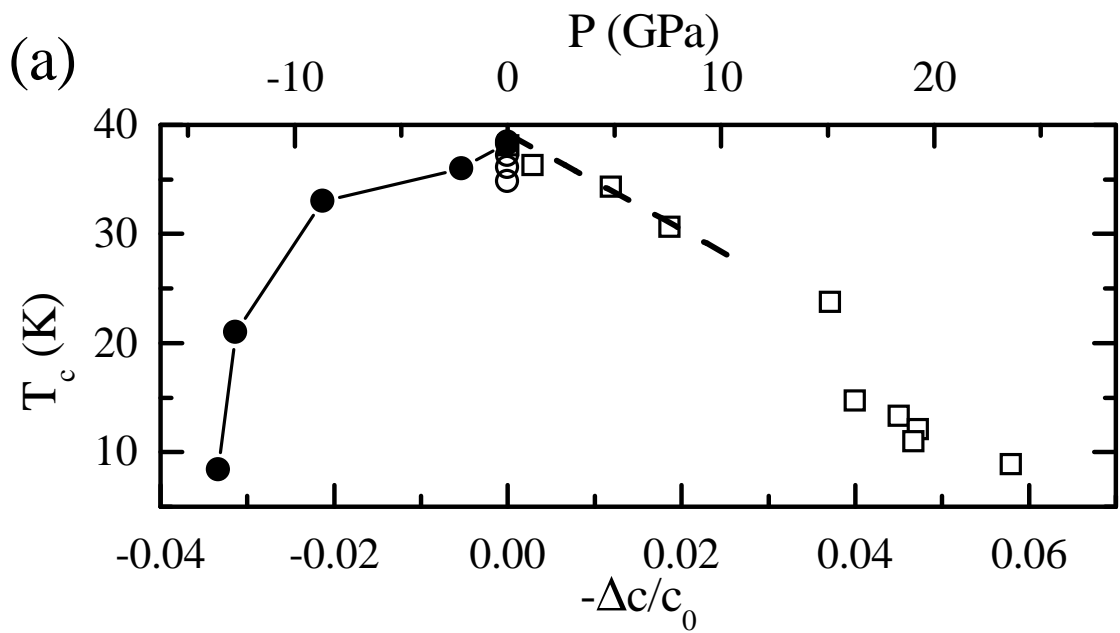
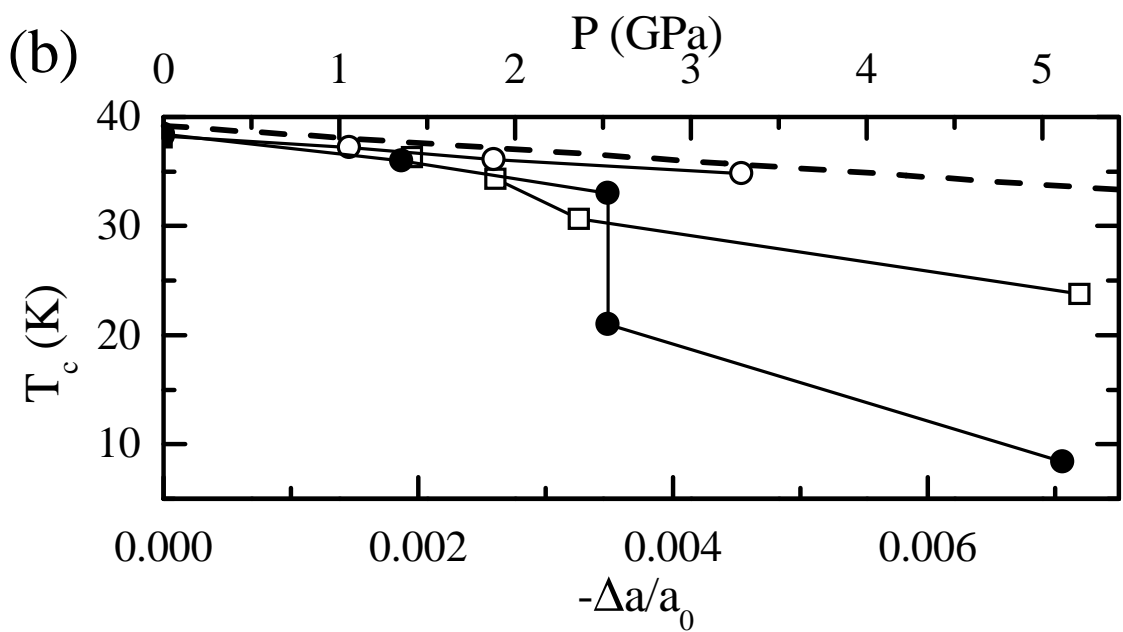
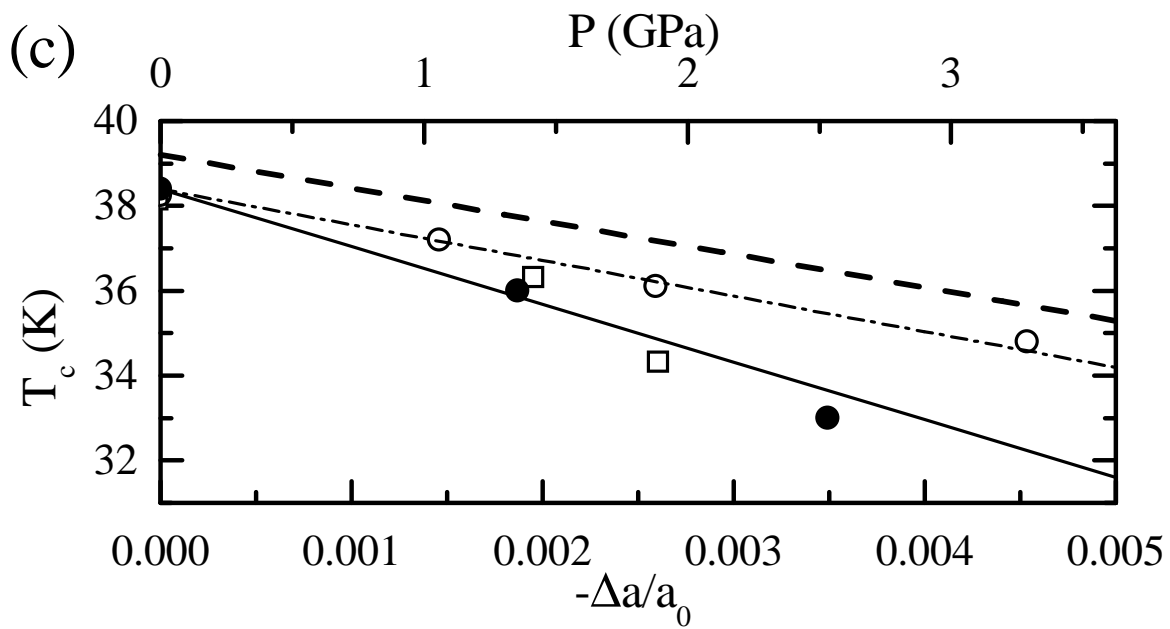
\end{document}